\documentclass[conference]{IEEEtran} 
\usepackage[utf8]{inputenc}
\usepackage{graphicx}
\usepackage{amsmath}
\usepackage{amsfonts}
\usepackage{amssymb}
\usepackage{xcolor}
\usepackage{listings} 
\usepackage{url} 
\usepackage[numbers]{natbib}
\usepackage{hyperref} 

\hypersetup{
    colorlinks=true,
    linkcolor=black, 
    citecolor=black,
    filecolor=black,
    urlcolor=black,
    pdftitle={ETDI: A Robust Security Layer for the Model Context Protocol using OAuth-Enhanced Tool Definitions},
    pdfauthor={Your Name(s)},
    pdfsubject={Computer Security, Large Language Models},
    pdfkeywords={MCP, LLM, Security, OAuth, Tool Poisoning, Rug Pull, Policy Engine, Access Control, Cedar, OPA}
}

\usepackage{academicons} 
\usepackage{xcolor} 

\newcommand{\orcidicon}[1]{%
    \href{https://orcid.org/#1}{%
        \includegraphics[width=10pt]{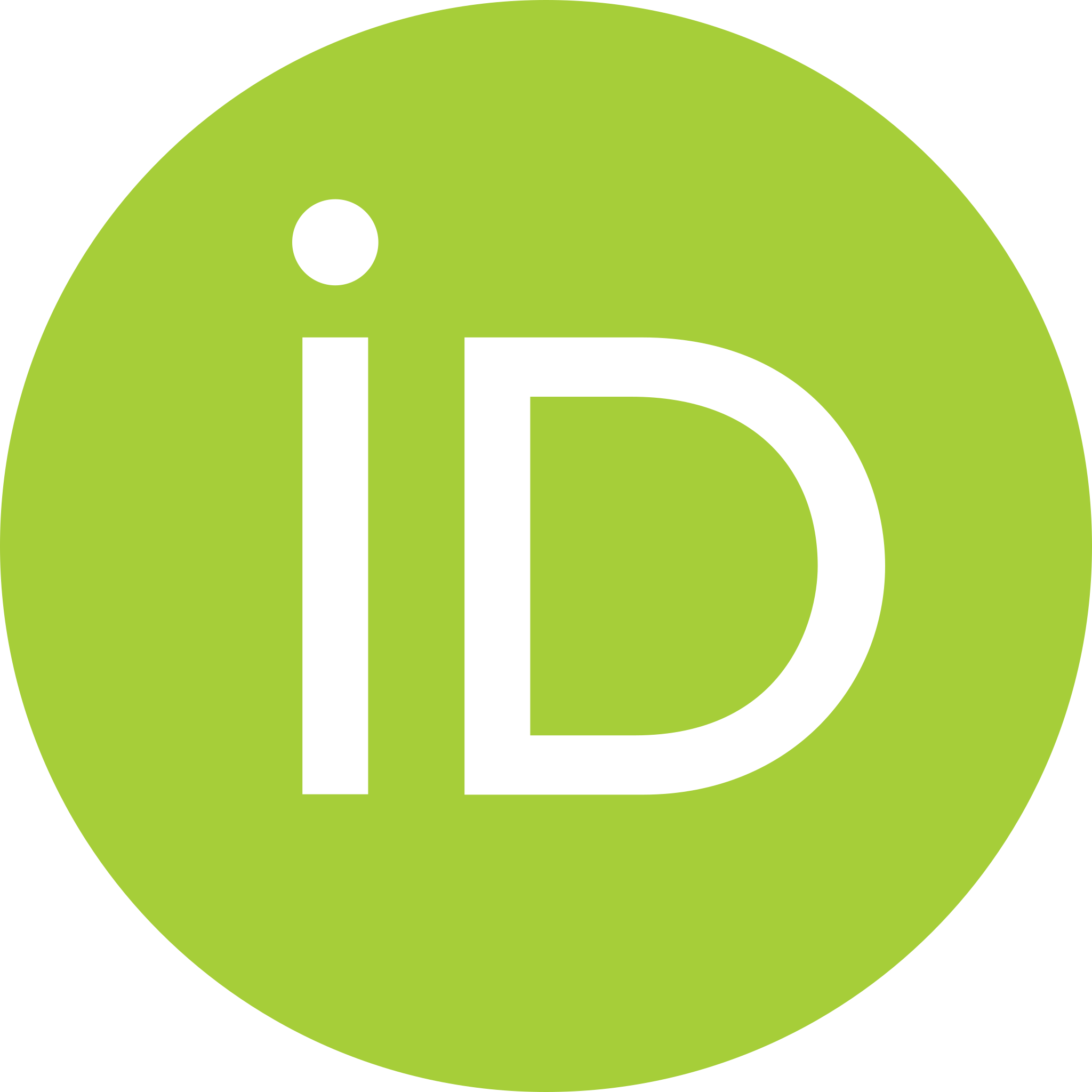}
    }%
}

\definecolor{codegreen}{rgb}{0,0.6,0}
\definecolor{codegray}{rgb}{0.5,0.5,0.5}
\definecolor{codepurple}{rgb}{0.58,0,0.82}
\definecolor{backcolour}{rgb}{0.95,0.95,0.92}

\lstdefinestyle{mystyle}{
    backgroundcolor=\color{backcolour},
    commentstyle=\color{codegreen},
    keywordstyle=\color{magenta}, 
    numberstyle=\tiny\color{codegray},
    stringstyle=\color{codepurple},
    basicstyle=\ttfamily\footnotesize,
    breakatwhitespace=true, 
    breaklines=true,
    captionpos=b,
    keepspaces=true,
    numbers=left,
    numbersep=5pt,
    showspaces=false,
    showstringspaces=false,
    showtabs=false,
    tabsize=2,
    frame=single,
    framerule=0.4pt,
    rulesepcolor=\color{gray},
    keywords={%
        IF, THEN, ELSE, ENDIF, FOR, WHILE, DO, ENDDO, FUNCTION, RETURN, TRUE, FALSE, NULL, 
        String, Integer, Boolean, Object, Array, 
        PolicyStore, Principal, Action, Resource, Context, IsAuthorized, AccessDenied, 
        invokeTool, verifyToolIdentity, getApprovedDefinition, verifyDefinitionIntegrity, 
        getVerifiedToolDefinition, getUserTokenForTool, 
        compareVersions, compareScopes, promptUserForApproval, storeApproval, log, 
        determineToolAction, currentTime 
    },
    morekeywords=[2]{
        HostApp, MCPClient, MCPServer, LLM, ToolDefinition, OAuthToken, JWT, IdP, User, 
        AmazonVerifiedPermissions, CedarPolicy, OpenPolicyAgent, Crypto 
    },
    keywordstyle=[1]\color{blue!70!black}\bfseries, 
    keywordstyle=[2]\color{green!40!black}\bfseries 
}
\title{ETDI: Mitigating Tool Squatting and Rug Pull Attacks in Model Context Protocol (MCP) by using OAuth-Enhanced Tool Definitions and Policy-Based Access Control}

\author{
\IEEEauthorblockN{Manish Bhatt\textsuperscript{1} \thanks{\textsuperscript{1}This work is not related to the author’s position at Amazon}}
\IEEEauthorblockA{\textit{Independent Researcher} \\
\textit{OWASP, Project Kuiper Security (KPES)} \\
Manish.bhatt13212@gmail.com \orcidicon{0000-0003-2207-5604} }
\and
\IEEEauthorblockN{Vineeth Sai Narajala\textsuperscript{2} \thanks{\textsuperscript{2}This work is not related to the author’s position at Amazon Web Services.}}
\IEEEauthorblockA{\textit{Proactive Security} \\
\textit{OWASP, Amazon Web Services} \\
vineeth.sai@owasp.org \orcidicon{0009-0007-4553-9930}}
\and
\IEEEauthorblockN{Idan Habler\textsuperscript{3}  \thanks{\textsuperscript{3}This work is not related to the author’s position at Intuit}}
\IEEEauthorblockA{\textit{Adversarial AI Security reSearch (A2RS)} \\
\textit{Intuit} \\
idan\_habler@intuit.com \orcidicon{0000-0003-3423-5927}}
}

\begin{document}

\maketitle

\begin{abstract}
The Model Context Protocol (MCP) plays a crucial role in extending the capabilities of Large Language Models (LLMs) by enabling integration with external tools and data sources. However, the standard MCP specification presents significant security vulnerabilities, notably Tool Poisoning and Rug Pull attacks. This paper introduces the Enhanced Tool Definition Interface (ETDI), a security extension designed to fortify MCP. ETDI incorporates cryptographic identity verification, immutable versioned tool definitions, and explicit permission management, often leveraging OAuth 2.0. We further propose extending MCP with fine-grained, policy-based access control, where tool capabilities are dynamically evaluated against explicit policies using a dedicated policy engine, considering runtime context beyond static OAuth scopes. This layered approach aims to establish a more secure, trustworthy, and controllable ecosystem for AI applications interacting with LLMs and external tools.
\end{abstract}

\let\thefootnote\relax\footnotetext{%
\textsuperscript{1}This work is not related to the author's position at Amazon. Co-equal Primary Author.

\textsuperscript{2}This work is not related to the author's position at Amazon Web Services. Co-equal Primary Author.

\textsuperscript{3}This work is not related to the author's position at Intuit.}


\section{Introduction}
The proliferation of Large Language Models (LLMs) marks a paradigm shift in human-computer interaction, with these models being embedded into a myriad of applications requiring nuanced understanding and generation of natural language. The Model Context Protocol (MCP) \cite{mcp_official_spec} has been instrumental in this evolution, providing a standardized mechanism for LLMs to interface with external tools, APIs, and data repositories. This contextual enrichment allows LLMs to transcend their inherent knowledge cut-offs, perform real-world actions, and offer dynamic, personalized user experiences.

Despite its functional benefits, the inherent openness and extensibility of the current MCP specification introduce significant security risks if not adequately protected \cite{reversinglabs_mcp_risks, willison_mcp_security}. The current MCP specification, while fostering innovation, lacks comprehensive security primitives to ensure the authenticity, integrity, and consistent behavior of integrated tools. This deficiency exposes users and systems to significant risks, primarily focusing on two critical attack vectors: Tool Poisoning and Rug Pull Attacks \cite{sarig_agent_threat_model}.

This paper introduces the Enhanced Tool Definition Interface (ETDI), a comprehensive security extension for MCP. ETDI's core is built on: (a) \textbf{Cryptographic Identity and Authenticity} for tools; (b) \textbf{Immutable and Versioned Definitions} to track changes; and (c) \textbf{Explicit and Verifiable Permissions}, often mapped to OAuth 2.0 scopes \cite{rfc6749} and conveyed via signed JSON Web Tokens (JWTs) \cite{rfc7519, rfc7515}.

Furthermore, we explore a significant enhancement: integrating \textbf{fine-grained, policy-based access control}. This involves extending MCP to support dynamic evaluation of tool actions by a policy engine (e.g., leveraging Amazon Verified Permissions with its Cedar policy language \cite{amazon_cedar_fse24}, or Open Policy Agent (OPA) \cite{opa_docs}). This allows defining what a tool can access based not just on its identity and static OAuth scopes, but also on the current context (time, location, user attributes, previous actions), enabling precise control over tool operations and supporting permission delegation policies.

This paper provides an in-depth analysis of MCP's architecture and vulnerabilities, a detailed exposition of ETDI, its OAuth-enhancement, the proposed policy-based access control extension, and a security analysis of these combined measures. We have demonstrated the feasibility of these enhancements here: \url{https://github.com/vineethsai/python-sdk} and detailed documentation here: \url{https://github.com/vineethsai/MCP-ETDI-docs}.

\section{The Model Context Protocol (MCP)}
Before we dive in, we need to understand how Standard MCP works. MCP \cite{mcp_official_spec} operates on a distributed client-server model, enabling LLM applications (Hosts) to connect via MCP Clients to MCP Servers that expose tools, resources, and prompts.

\subsection{Architecture Overview}
\begin{figure}[h!]
  \centering
  \includegraphics[width=1\columnwidth]{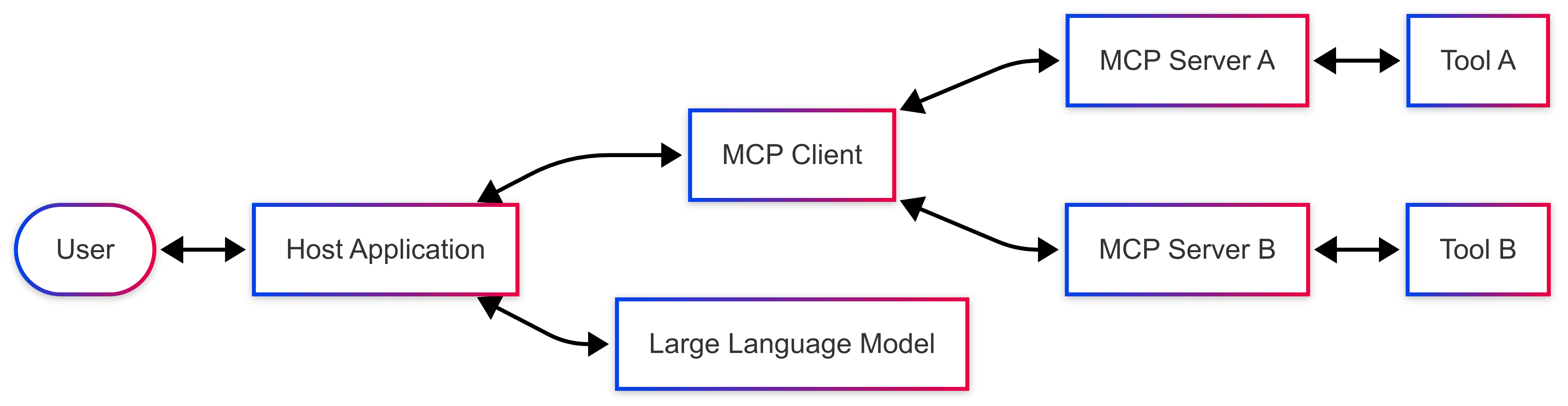} 
  \caption{High-Level MCP Architecture - This diagram shows the interaction flow between the user, host application, MCP client, LLM, and various MCP servers providing tools.}
  \label{fig:mcp_architecture}
\end{figure}

MCP operates on a distributed client-server model. Key components include:
\begin{itemize}
    \item \textbf{Host Applications}: User-facing applications (e.g., AI-powered desktop apps, IDE extensions) that orchestrate interactions.
    \item \textbf{MCP Clients}: Software components within Host Applications that discover, connect to, and interact with MCP Servers.
    \item \textbf{MCP Servers}: Services that expose capabilities (tools, resources, prompts) to MCP Clients.
    \item \textbf{Tools}: Discrete functions or services invokable by an LLM via an MCP Server.
    \item \textbf{Resources}: Data sources accessible by the LLM for context.
    \item \textbf{Prompts}: Pre-defined templates guiding LLM tool/resource usage.
\end{itemize}
Figure \ref{fig:mcp_architecture} illustrates the high-level MCP architecture.

\subsection{Standard Operational Flow}
The MCP interaction involves two main phases:

\subsubsection{Initialization and Discovery Phase}
\begin{figure}[h!]
  \centering
  \includegraphics[width=1\columnwidth]{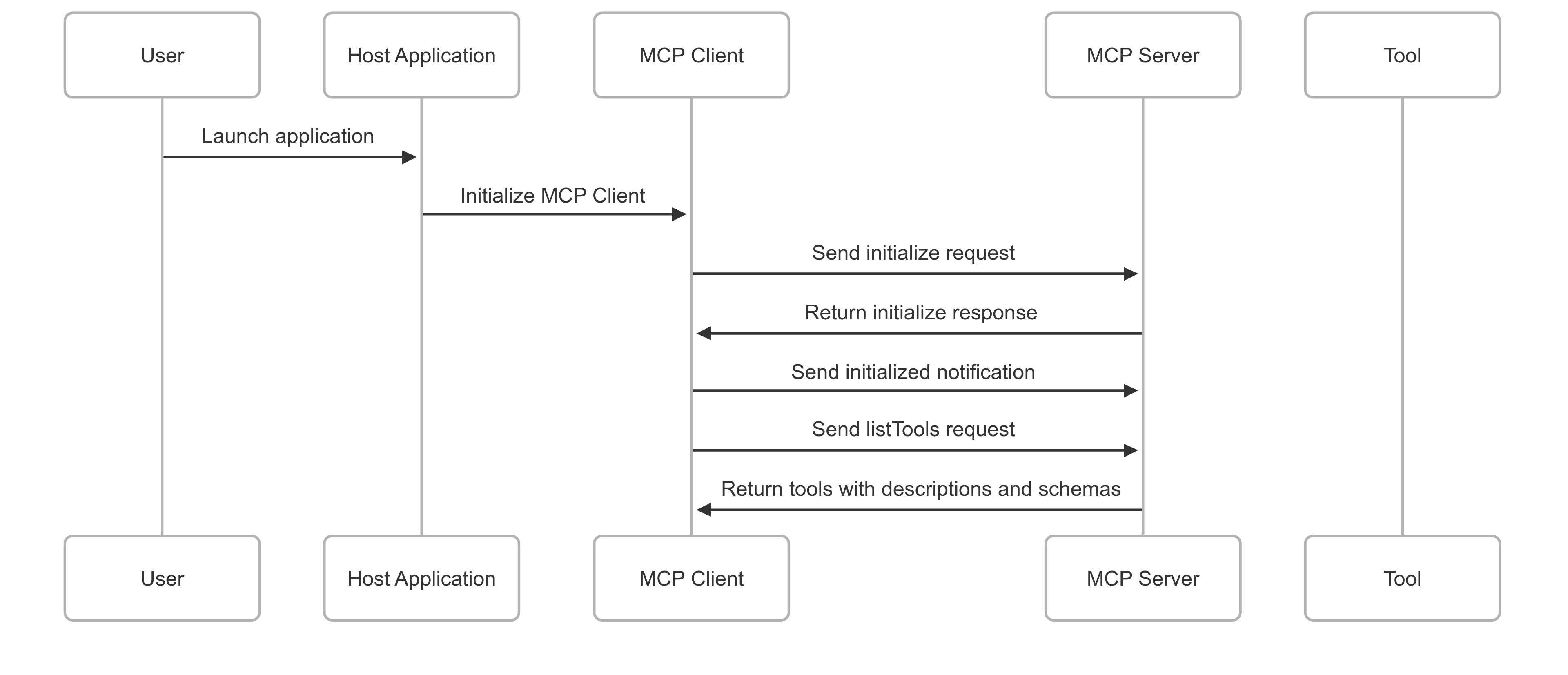} 
  \caption{MCP Initialization and Tool Discovery Sequence.}
  \label{fig:init-and-discovery}
\end{figure}

This phase establishes the operational context between MCP Clients and Servers.
\begin{enumerate}
    \item \textbf{Application Launch \& Client Initialization}: Upon startup, the Host Application initializes its embedded MCP Client module(s). This client is responsible for managing all subsequent MCP communications.
    \item \textbf{Server Handshake and Capability Negotiation}: MCP Clients initiate a handshake sequence (e.g., an initialize request) with known or discoverable MCP Servers. During this handshake, servers may declare their supported protocol versions, specific capabilities (including ETDI support if present), and other relevant metadata. Clients, in turn, may also communicate their capabilities.
    \item \textbf{Tool Listing Request}: Once the connection is established and capabilities are negotiated, the MCP Client typically sends a listTools (or equivalent) request to each connected MCP Server to enumerate the available tools.
    \item \textbf{Tool Definition Exchange}: MCP Servers respond with a list of tool definitions. Crucially, in the standard MCP, these definitions include human-readable descriptions, a machine-readable name or ID, and a JSON schema detailing expected input parameters and output format. However, they lack verifiable authenticity or integrity markers.
\end{enumerate}
An end-to-end flow of the standard operational flow, including initialization and tool discovery phase is illustrated in Figure~\ref{fig:init-and-discovery}

\subsubsection{Tool Invocation and Usage Phase}
\begin{figure}[h!]
  \centering
  \includegraphics[width=1\columnwidth]{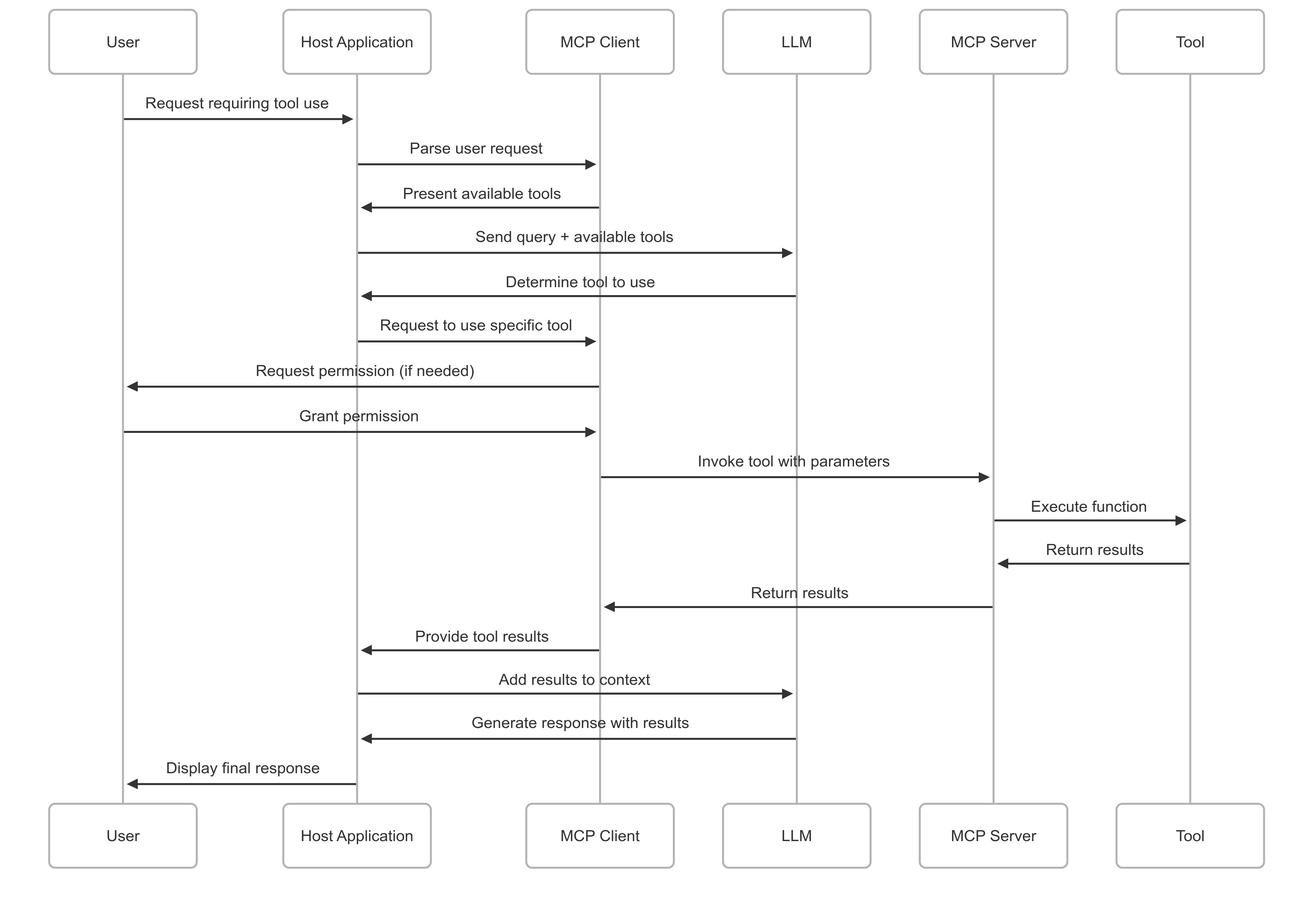} 
  \caption{MCP Tool Usage and Invocation Sequence.}
  \label{fig:tool-invocation-and-usage}
\end{figure}

This phase describes the dynamic interaction leading to tool execution.
\begin{enumerate}
    \item \textbf{User Request Processing}: The user interacts with the Host Application, issuing a request or command that may benefit from external tool capabilities (e.g., "What's the weather in London?", "Summarize the key points of document.pdf").
    \item \textbf{Tool Selection by LLM/Host Application}: The Host Application, often in conjunction with an LLM, processes the user's request. The LLM, having been provided with the (unverified in standard MCP) descriptions and schemas of available tools, identifies a suitable tool and determines the necessary input parameters based on the user's query.
    \item \textbf{Permission Adjudication (Conditional and Flawed)}: If the selected tool is flagged as requiring specific permissions (e.g., access to location services, file system, or incurring costs), or if it's the first time a user encounters this tool (by name/ID), the MCP Client (via the Host Application) might prompt the user for explicit approval. However, this approval is based on potentially spoofed or misleading tool descriptions.
    \item \textbf{Tool Invocation Command}: Upon receiving approval (if sought), the MCP Client dispatches an invokeTool (or similar) command to the relevant MCP Server, specifying the tool ID and the parameters derived by the LLM.
    \item \textbf{Server-Side Tool Execution}: The MCP Server receives the invocation command and delegates the request to the actual tool implementation. The tool then executes its defined function, which could involve interacting with other services, performing computations, or accessing data.
    \item \textbf{Result Propagation}: The tool returns its output (or an error status) to the MCP Server, which then relays this information back to the MCP Client that initiated the request.
    \item \textbf{Context Augmentation and LLM Response Generation}: The MCP Client provides the tool's results to the Host Application. These results are typically incorporated into the LLM's context. The LLM then uses this newly augmented context to generate a final, more informed, and potentially action-oriented response to the user's original query.
\end{enumerate}
An example of tool usage and invocation flow  is illustrated in Figure~\ref{fig:tool-invocation-and-usage}.

\section{Critical Security Vulnerabilities in MCP}
The standard MCP flow, lacking robust tool identity and integrity verification, is susceptible to critical attacks \cite{reversinglabs_mcp_risks}.

\subsection{Attack Vector 1: Tool Poisoning}
\textbf{Definition}: Tool Poisoning involves a malicious actor deploying a tool that masquerades as a legitimate, trusted, or innocuous tool \cite{willison_mcp_security}. The attacker's objective is to deceive either the end-user or the LLM (during automated tool selection) into selecting and approving the malicious tool, thereby granting it unauthorized access or capabilities.

\begin{figure}[h!]
  \centering
  \includegraphics[width=1\columnwidth]{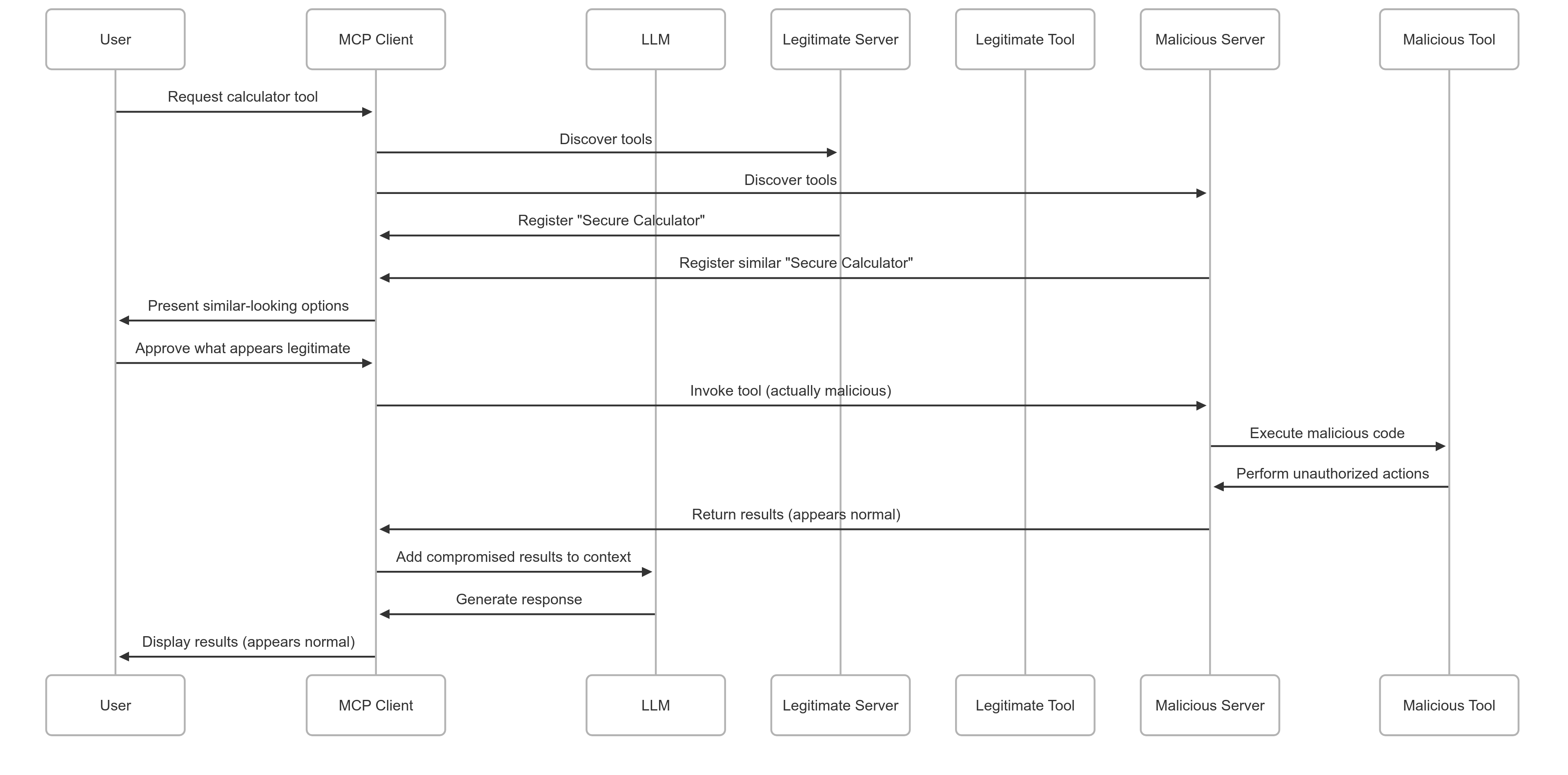} 
  \caption{Tool Poisoning Attack Sequence.}
  \label{fig:tool-poisoning}
\end{figure}

\textbf{Vulnerability Analysis}: The susceptibility to Tool Poisoning in standard MCP stems from several deficiencies:
\begin{itemize}
    \item \textbf{Lack of Authenticity Verification}: There is no built-in mechanism for MCP Clients or users to cryptographically verify the true origin or authenticity of a tool. Tool names, descriptions, and even claimed provider names within the tool's metadata can be easily spoofed.
    \item \textbf{Indistinguishable Duplicates and Ambiguity}: If a malicious tool meticulously replicates the metadata (name, description, JSON schema) of a legitimate tool, it becomes virtually impossible for the user or an automated LLM-based selection process to differentiate between them. This is especially problematic if multiple tools with similar names appear, and the selection criteria are not robust.
    \item \textbf{Exploitation of Implicit Trust}: Attackers leverage the user's inherent trust in familiar tool names (e.g., "Calculator," "Calendar Access") or reputable provider names. A tool might falsely claim to be from "TrustedSoft Inc." in its description.
    \item \textbf{Unverifiable Claims in Descriptions}: A tool can assert claims like "secure," "official," or "privacy-preserving" in its human-readable description without any underlying mechanism to validate these assertions.
\end{itemize}
\textbf{Impact}: Successful tool poisoning (as illustrated in Figure~\ref{fig:tool-poisoning} can lead to a wide array of detrimental consequences, including but not limited to: exfiltration of sensitive personal or corporate data, unauthorized execution of system commands, installation of malware or ransomware, financial fraud through manipulated transactions, and the subtle manipulation of LLM outputs to spread misinformation or achieve other nefarious goals.

\textbf{Illustrative Scenario}: An attacker deploys a malicious MCP server hosting a tool named "SecureDocs Scanner." They meticulously copy the description, JSON schema, and even claim "TrustedSoft Inc." as the provider in the tool's metadata. The user's MCP Client discovers both the legitimate and the malicious "SecureDocs Scanner" tools. Due to identical presentation, they appear as duplicates, or the client might even de-duplicate them, potentially favoring the malicious one based on arbitrary factors like discovery order. The user, intending to use the trusted tool, selects the entry that corresponds to the malicious version, or the LLM selects it. Upon invocation, the malicious "SecureDocs Scanner" exfiltrates the entire content of any document processed through it to an attacker-controlled server, while possibly returning a fake "No PII found" message.

\subsection{Attack Vector 2: Rug Pull Attacks}
\begin{figure}[h!]
  \centering
  \includegraphics[width=1\columnwidth]{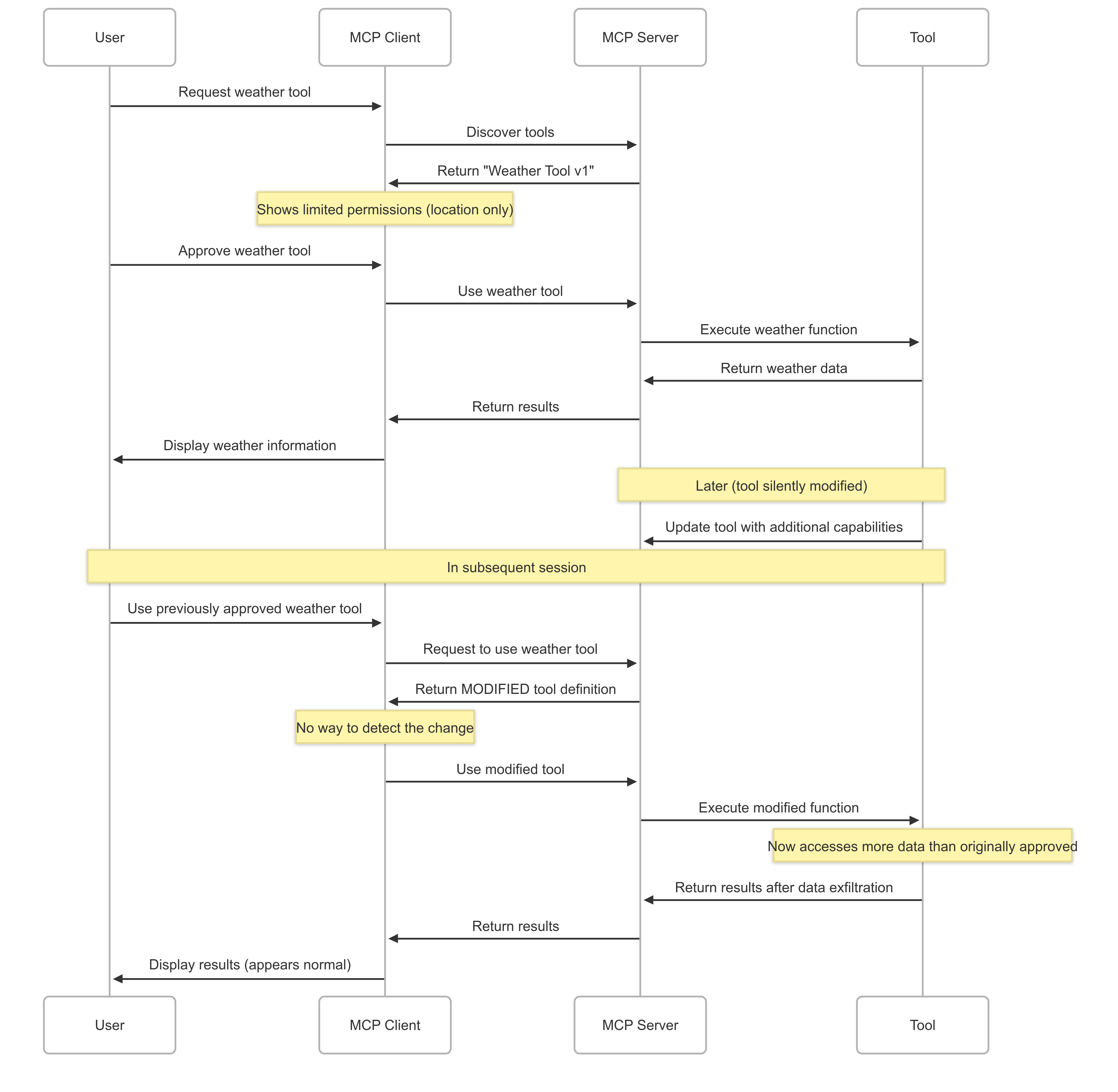} 
  \caption{Rug Pull Attack Sequence.}
  \label{fig:rug-pull}
\end{figure}

\textbf{Definition}: Rug Pull attacks, also known as "bait-and-switch" in this context, manifest when the functionality, data access patterns, or permission requirements of an already approved tool are maliciously altered by its provider \textbf{after} the initial user consent has been granted \cite{willison_mcp_security}. The tool initially presents benign or expected behavior to gain trust and approval, then later changes to perform unauthorized actions without re-triggering a consent request.
An example of Rug Pull attack is illustrated in Figure~\ref{fig:rug-pull}

\textbf{Vulnerability Analysis}: The core vulnerabilities enabling Rug Pulls are:
\begin{itemize}
    \item \textbf{Mutability of Server-Side Logic}: The fundamental issue is that a tool's underlying code and behavior on the MCP Server can be modified without any notification to, or re-verification by, the MCP Client or user, especially if the tool's primary identifier (e.g., name) remains static.
    \item \textbf{Lack of Continuous Integrity Checks}: Standard MCP Clients, once a tool is "approved" (often based on its name or a transient session approval), typically do not re-fetch and re-verify the tool's complete definition (including its schema or a cryptographic hash) on every subsequent invocation.
    \item \textbf{Absence of Re-Approval Triggers for Definition Changes}: If the tool's identifier (like its name or version string, if superficially unchanged) doesn't change, or if the client isn't designed to detect subtle modifications in its JSON schema or descriptive metadata, no new approval prompt is presented to the user. The client might continue to operate under the assumption that the tool's behavior aligns with the initially approved state.
    \item \textbf{Exploitation of Established Trust}: The attack leverages the trust established during the initial, benign approval phase. Users are unlikely to scrutinize a tool they believe they have already vetted.
\end{itemize}
\textbf{Impact}: Rug Pulls can lead to severe breaches of privacy and security, such as unauthorized access to sensitive data (e.g., private conversations, files, contact lists, financial information) that the user never explicitly consented to share with the *modified* version of the tool. This effectively bypasses the initial permission model and can lead to a profound loss of user trust once discovered.

\textbf{Illustrative Scenario}: A user installs and approves a "Daily Wallpaper" tool. Version 1.0 of this tool fetches a new wallpaper image and sets it. It requests permission only to "access the internet" and "modify desktop wallpaper." Weeks later, the provider (or an attacker who compromised the server) updates the tool's server-side logic. The tool, still identified as "Daily Wallpaper v1.0" to avoid re-approval, is now modified to also scan the user's Documents folder for files containing financial keywords and upload them. The next time the tool runs, it performs this malicious action silently in the background.

\section{ETDI: The Enhanced Tool Definition Interface}
ETDI is a security layer extension for MCP designed to counter Tool Poisoning and Rug Pulls by introducing verifiable identity and integrity for tool definitions.

\subsection{Foundational Security Principles of ETDI}
ETDI is architected upon three fundamental security principles:
\begin{enumerate}
    \item \textbf{Cryptographic Identity and Authenticity}: Tool definitions are digitally signed by the provider. MCP Clients verify these signatures.
    \item \textbf{Immutable and Versioned Definitions}: Any change to a tool's definition (functionality, metadata, schema, permissions, or a hash of its backend API contract like an OpenAPI specification) mandates a new, signed version. This helps detect unauthorized modifications and API contract drift.
    \item \textbf{Explicit and Verifiable Permissions}: A tool's required capabilities (e.g., OAuth scopes) are declared in its signed definition. Critically, the tool definition should also declare any specific permissions or entitlements the *user or calling application* is expected to possess and present to the tool for it to operate (e.g., subscription level for a paid service). These are presented to the user for approval.
\end{enumerate}

A simplified pseudo-code for ETDI client-side verification of a tool definition upon first encounter or update might look like the listing in listing ~\ref{lst:etdi_verification}.

\begin{lstlisting}[caption={Simplified ETDI Tool Definition Verification Logic}, label={lst:etdi_verification}]
FUNCTION verifyAndApproveTool(toolDefinition, providerPublicKey)
  // 1. Verify cryptographic identity and integrity
  IF NOT Crypto.verifySignature(toolDefinition.content, toolDefinition.signature, providerPublicKey) THEN
    log("Tool definition signature invalid for " + toolDefinition.id);
    RETURN FALSE; // Reject tool
  ENDIF

  // 2. Check against previously approved version (if any)
  approvedDef = getApprovedDefinition(toolDefinition.id);
  IF approvedDef != NULL THEN
    IF toolDefinition.version == approvedDef.version THEN
      // If version is same, ensure definition content hash matches stored hash
      IF Crypto.hash(toolDefinition.content) != approvedDef.contentHash THEN
        log("Tampering detected for " + toolDefinition.id + " version " + toolDefinition.version);
        // Requires re-approval even for same version if content changed
        IF NOT promptUserForApproval("Tool " + toolDefinition.id + " content changed. Re-approve?", toolDefinition) THEN
            RETURN FALSE;
        ENDIF
      ENDIF
    ELSE IF toolDefinition.version < approvedDef.version THEN
      log("Warning: Older version " + toolDefinition.version + " of tool " + toolDefinition.id + " presented.");
      // Policy decision: allow, warn, or block older versions
    ELSE // New version
      IF NOT promptUserForApproval("New version " + toolDefinition.version + " for tool " + toolDefinition.id + ". Approve?", toolDefinition) THEN
        RETURN FALSE;
      ENDIF
    ENDIF
  ELSE // First time seeing this tool
    IF NOT promptUserForApproval("Approve new tool " + toolDefinition.id + "?", toolDefinition) THEN
      RETURN FALSE;
    ENDIF
  ENDIF
  
  // 3. Store/update approval with new definition hash and version
  storeApproval(toolDefinition.id, toolDefinition.version, Crypto.hash(toolDefinition.content), toolDefinition.permissions);
  RETURN TRUE;
ENDFUNCTION
\end{lstlisting}

\subsection{ETDI Countermeasures}
\subsubsection{Thwarting Tool Poisoning}
\begin{figure}[t!]
  \centering
  \includegraphics[width=1\columnwidth]{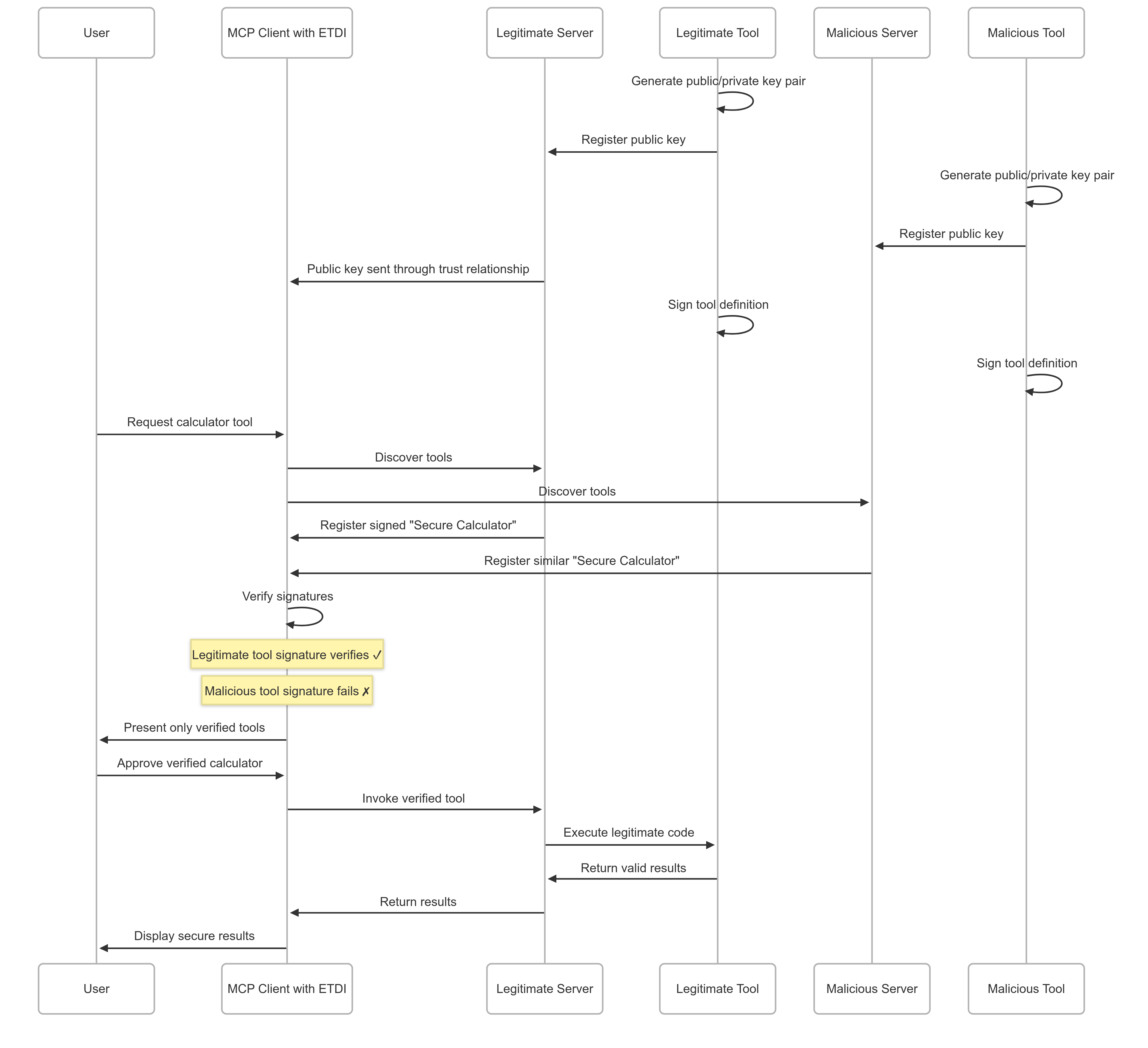} 
  \caption{ETDI Preventing Tool Poisoning through Cryptographic Signatures.}
  \label{fig:prevent-tool-poisoning}
\end{figure}

ETDI's design directly counters Tool Poisoning by establishing a chain of trust rooted in cryptographic verification:
\begin{itemize}
    \item \textbf{Provider Key Infrastructure}: Legitimate tool providers generate a public/private cryptographic key pair (e.g., RSA, ECDSA). The private key is kept secret by the provider, while the public key is made securely available to MCP Clients. This can be achieved through various mechanisms, such as distribution with the Host Application, a trusted key registry, or via a PKI.
    \item \textbf{Digitally Signed Tool Definitions}: When a provider creates or updates a tool, they generate a comprehensive tool definition encompassing its name, description, input/output JSON schema, semantic version, and a detailed list of required permissions. This entire definition is then digitally signed using the provider's private key, producing a signature that is bundled with the definition.
    \item \textbf{Mandatory Client-Side Verification}: When an ETDI-enabled MCP Client discovers tools, it receives these signed definitions. Before presenting a tool to the user or LLM, the client \textit{must} verify the digital signature using the claimed provider's public key. This verification confirms that the definition was indeed signed by the legitimate provider and has not been altered in transit.
    \item \textbf{Policy Enforcement for Unverified Tools}: If a signature is invalid (e.g., fails verification, signed by an untrusted key) or missing, the tool is flagged as unverified or potentially malicious. The ETDI client can then enforce policies such as hiding such tools from the user, displaying prominent warnings, or entirely preventing their invocation.
\end{itemize}
Consequently, an attacker cannot forge a valid signature for a tool they don't legitimately own unless they compromise the legitimate provider's private key. This makes simple impersonation computationally infeasible.
Figure~\ref{fig:prevent-tool-poisoning} demonstrate how ETDI can be used to prevent tool poisoning attack through cryptographic signatures.

\subsubsection{Preventing Rug Pulls}
ETDI's immutability and versioning principles are key. An important aspect is that if the tool definition includes a hash of its backend API contract (e.g., derived from an OpenAPI/Swagger specification), any change to this contract by the tool provider (even if the tool's descriptive metadata remains the same) would alter the hash. This change would necessitate a new tool definition version, triggering re-approval by the user and making such backend changes transparent.

\begin{figure}[h!]
  \centering
  \includegraphics[width=1\columnwidth]{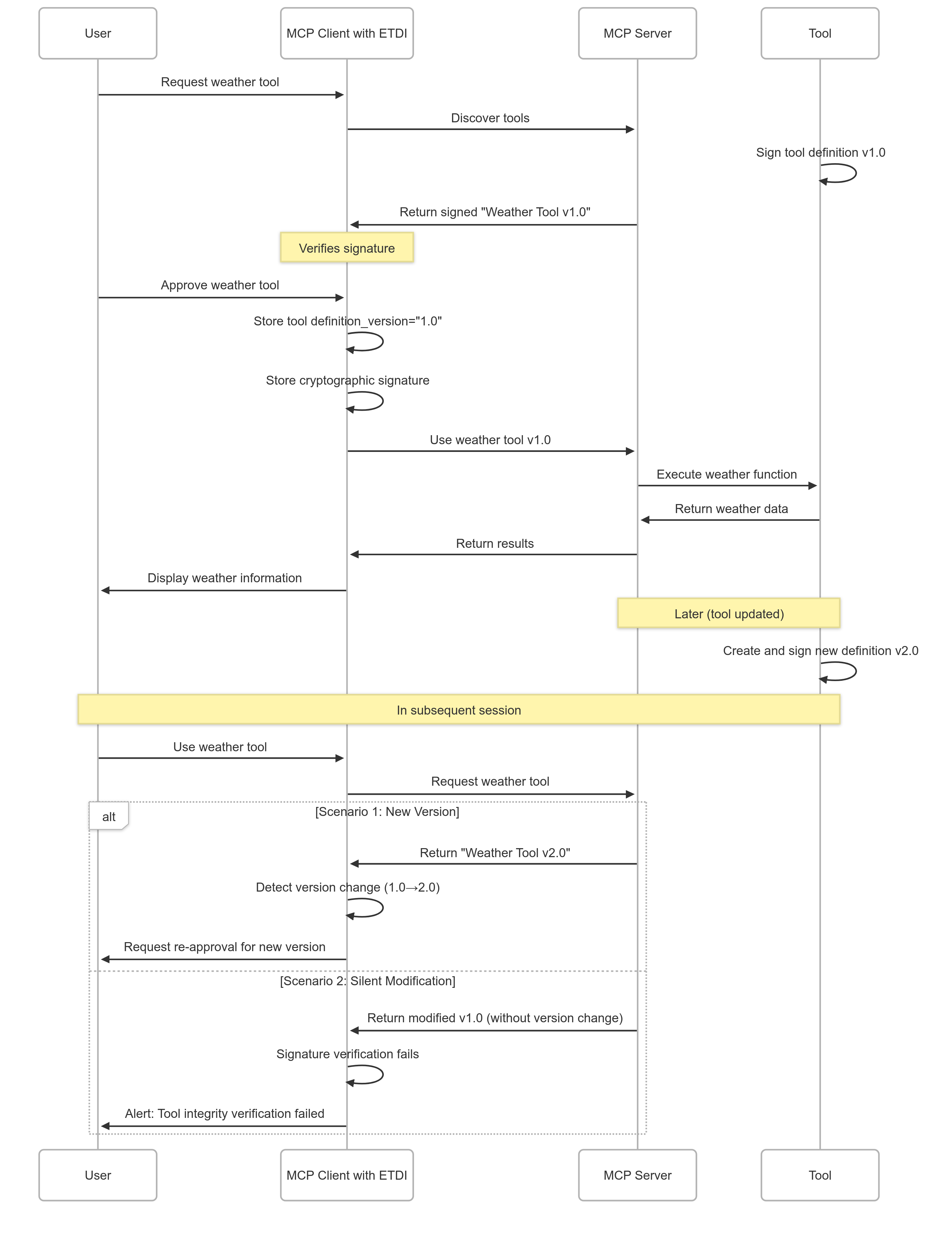} 
  \caption{ETDI Preventing Rug Pulls through Versioning and Signature Verification.}
  \label{fig:prevent-rug-pull}
\end{figure}

Figure~\ref{fig:prevent-rug-pull} demonstrate how ETDI can be used to prevent rug pull attacks through versioning and signature verification.

\subsection{Advancing Security with OAuth-Enhanced ETDI}
While direct cryptographic signatures provide a strong security foundation, integrating ETDI with an established authorization framework like OAuth 2.0 \cite{rfc6749} offers significant advantages in terms of standardization, ecosystem interoperability, fine-grained access control, and centralized trust management.

\begin{figure}[h!]
  \centering
  \includegraphics[width=1\columnwidth]{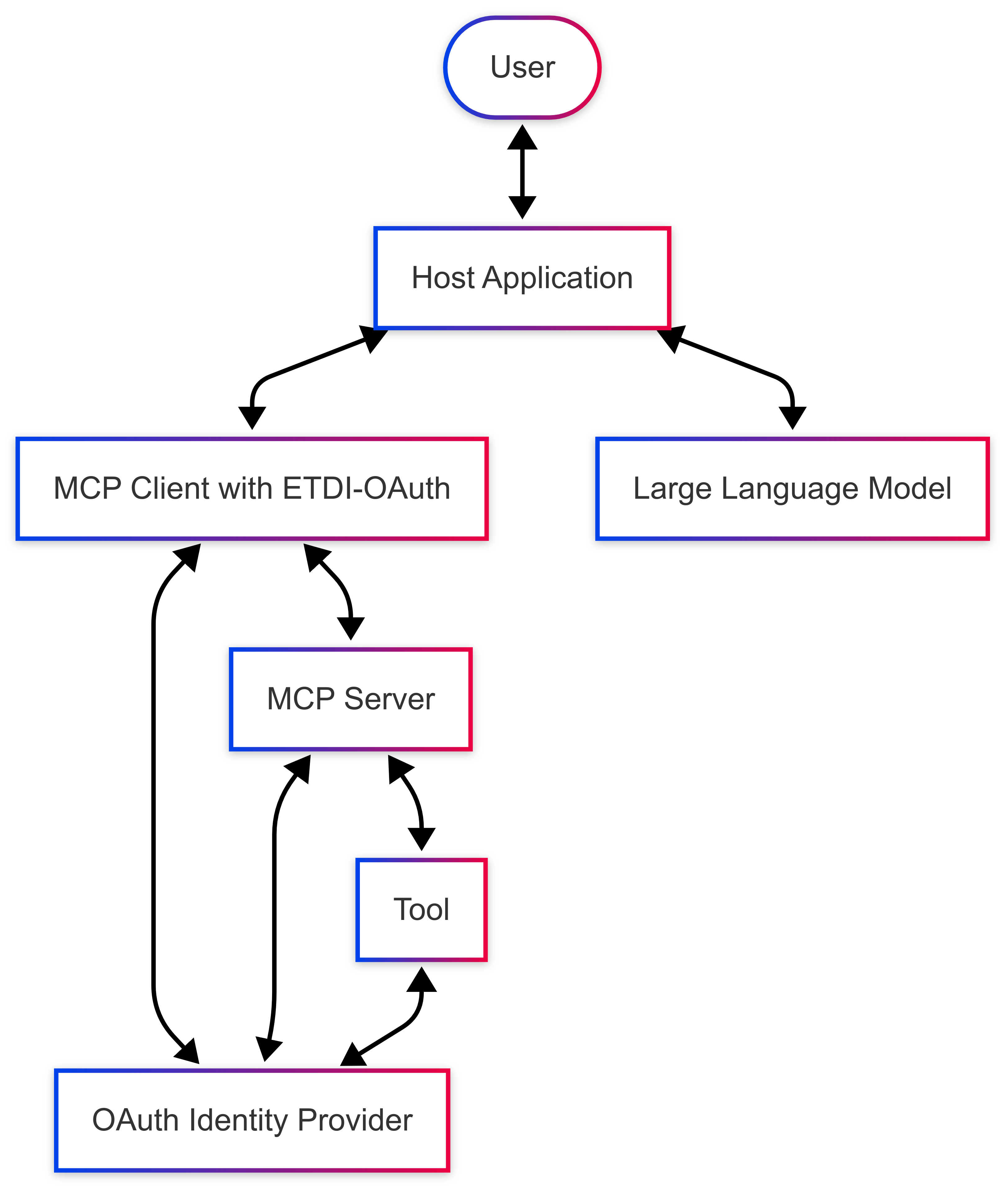} 
  \caption{OAuth-Enhanced ETDI Architecture, introducing an Identity Provider.}
  \label{fig:oauth2-integ}
\end{figure}

It is also important to distinguish between ETDI verifying the tool's identity and permissions to the MCP client/user, and the tool itself potentially needing to verify the *user's or calling application's authorization* to use its services. In many real-world scenarios, particularly for commercial tools or APIs, the tool acts as an OAuth Resource Server. The MCP Host Application (acting as an OAuth Client) would obtain an OAuth token for the user (e.g., an access token from an IdP like Google, or a custom token indicating user entitlements like "premium\_access") and pass this token with the invocation request to the MCP Server, which then forwards it to the tool. The tool would validate this token to ensure the user/application is authorized for the requested operation (e.g., has paid for the service). The tool's ETDI definition should declare what kind of user-level authentication/authorization (e.g., specific OAuth issuer, required scopes in the user's token) it expects.
The architecture of OAuth-Enhanced ETDI is illustrated in Figure~\ref{fig:oauth2-integ}

\subsection{EExample Workflow with Amaicy-Based Access Control}
While ETDI, especially with OAuth, provides robust tool identity and permission scoping, integrating fine-grained, context-aware access control using dedicated policy engines offers a powerful enhancement. This approach moves beyond static permission declarations to dynamic evaluation based on the runtime environment and detailed policies.

\subsubsection{Concept and Architecture}
MCP servers (or the host applications themselves) would integrate with a Policy Decision Point (PDP). This PDP could be an existing solution like Open Policy Agent (OPA) \cite{opa_docs} or a service like Amazon Verified Permissions, which utilizes the Cedar policy language \cite{amazon_cedar_fse24}. Tool definitions would be augmented to include or reference specific resource access control policies that detail what resources the tool can access under specific conditions (e.g., time of day, user location, nature of data, previous actions). These policies themselves could be signed artifacts managed within a Policy Administration Point (PAP).

When an MCP Client intends to invoke a tool, after the standard ETDI identity and integrity checks, a request is made to the PDP. This request would typically include:
\begin{itemize}
    \item \textbf{Principal}: The authenticated identity of the tool (e.g., derived from its ETDI-OAuth token).
    \item \textbf{Action}: The specific operation the tool intends to perform (e.g., `ReadFile`, `SendEmail`).
    \item \textbf{Resource}: The target resource of the action (e.g., `urn:filesystem:/user/docs/file.txt`, `urn:email:recipient@example.com`).
    \item \textbf{Context}: A rich set of contextual attributes, such as user identity, device security posture, time, location, purpose of the request, or data sensitivity.
\end{itemize}
The PDP evaluates these inputs against the applicable policies and returns a decision (Permit/Deny).

\subsubsection{Policy Server Interaction and Signed Policies}
The policy engine (PDP) and the policies it uses must be part of the trusted computing base.
\begin{itemize}
    \item \textbf{PDP Discovery and Authentication}: PDP endpoints could be discovered via a trusted service registry \cite{Narajala2025ToolSquatting} or be pre-configured in MCP clients/servers. Communication must be secured (e.g., using mTLS).
    \item \textbf{Policy Integrity and Distribution}: The policies themselves (e.g., Cedar policies) must be securely managed and distributed. If policies are treated as signed artifacts, the PDP (or a component feeding policies to it) would verify policy signatures before loading or using them, ensuring they originate from a trusted policy administrator and haven't been tampered with. This creates a chain of trust for policy definition and enforcement. The process of evaluating signed policies involves the PDP fetching the relevant policy, verifying its signature against a trusted key, and then executing the policy logic against the request attributes.
\end{itemize}
This model aligns with authorization patterns seen in microservices architectures, where centralized policy management and enforcement provide consistent security \cite{oso_microservices_auth}.

\subsubsection{Example Workflow with Amazon Verified Permissions}
The pseudo-code in listing~\ref{lst:avp_check_pseudo}  illustrates a workflow where an MCP system component consults Amazon Verified Permissions before allowing a tool invocation.


\begin{lstlisting}[caption={Tool Invocation with Amazon Verified Permissions Policy Check}, label={lst:avp_check_pseudo}]
// Context: LLM has determined ToolX is needed to access ResourceY for UserZ
// HostApp or MCPClient is orchestrating this.

FUNCTION invokeToolWithPolicyCheck(toolId, resourceId, userContext)
  // 1. Retrieve and verify ToolX's definition via ETDI (OAuth/signature)
  toolDefinition = MCPClient.getVerifiedToolDefinition(toolId);
  IF toolDefinition == NULL THEN
    RETURN AccessDenied("Tool identity/integrity verification failed for " + toolId);
  ENDIF

  // 2. Prepare policy evaluation request attributes
  principal = toolDefinition.identity; // e.g., "ToolVendor::ToolX_v1.2"
  action = determineToolAction(toolDefinition, resourceId); // e.g., "File::Read", "API::Invoke"
  resource = resourceId; // e.g., "UserDocs::Private::AnnualReport.pdf"
  
  // Enrich context with user and environment data
  context = {
    user: { id: userContext.userId, department: userContext.department },
    request: { time: currentTime(), purpose: userContext.statedPurpose },
    // Potentially, information about data sensitivity, location, etc.
  };

  // 3. Call Policy Engine (e.g., Amazon Verified Permissions)
  // policyStoreId identifies the set of applicable Cedar policies
  policyStoreId = "userZ_policy_store"; // Could be user-specific or group-specific
  
  authDecision = AmazonVerifiedPermissions.isAuthorized(
                      policyStoreId,
                      principal,
                      action,
                      resource,
                      context
                  );

  // 4. AVP evaluates applicable Cedar policies (internal to AVP service)

  // 5. If authorized, tool invocation proceeds via MCP Server
  IF authDecision.isAllowed() THEN
    log("Policy check passed for " + toolId + " on " + resourceId);
    // If tool requires user-specific token, HostApp retrieves and passes it
    userAuthToken = HostApp.getUserTokenForTool(toolId); 
    toolResult = MCPServer.invokeTool(toolDefinition, resourceId, userAuthToken);
    RETURN toolResult;
  ELSE
  // 6. Access denied if policies don't permit
    log("Policy check failed for " + toolId + ": " + authDecision.errors());
    RETURN AccessDenied("Tool " + toolId + " not authorized by policy for this action/resource/context.");
  ENDIF
ENDFUNCTION
\end{lstlisting}

\subsubsection{Policy-Based Call Stack Verification for Tool Chain Security}
    Fine-grained requirements are applicable not just to restrict tool access to particular resources but also to the dynamic chaining of tool invocations. This is accomplished using \textbf{Call Stack Verification}, an approach that keeps track of and enforces guidelines on the order of tool calls made during an active session. In this case, the series of tools that have called each other, leading to the current invocation, is represented by a call stack. The \textbf{callee} is the tool invoked and the \textbf{ caller} is the tool initiating the invocation.

    The main goals of Call Stack Verification are to prevent:
    \begin{itemize}  
        \item Enforcing rules that specify acceptable tool call sequences (e.g., Tool A can call Tool B, but Tool C cannot call Tool D directly) is denoted as "Unauthorized Tool Chaining." Allow and block lists for particular caller-callee pairings, as specified in a \texttt{CallStackPolicy}, can be used to manage chaining.
        \item \textbf{Privilege Escalation Through Tool Calls}: Identifying and preventing scenarios in which a caller tool with lower privileges calls a callee tool with higherprivileges in a manner that improperly raises effective permissions without explicit approval. This involves contrasting the caller's and callee's authorization scopes.
        \item \textbf{Circular Call Dependencies}: Unless specifically allowed by policy, identifying and stopping call sequences in which a tool calls itself directly or indirectly (e.g., A $\rightarrow$ B $\rightarrow$ A) can result in resource exhaustion or infinite loops.
        \item \textbf{Excessive Call Depth Attacks}: Preventing denial-of-service attempts that take advantage of deep call chains or stack overflow-like situations by limiting the maximum number of nested tool calls.
        \item \textbf{Rate Limit Violations}: Preventing misuse or resource exhaustion by imposing restrictions on the number of times a particular tool may be called in a given time frame.
    \end{itemize}
    
    The state of the active call stack for every session would be maintained by a specialized component, like a \texttt{ETDI Call Stack Verifier}. This verifier checks the call against the active \texttt{CallStackPolicy} on each attempt to invoke the tool. This policy establishes guidelines for circular calls, allowed/blocked chains, maximum call depth, and criteria for identifying privilege escalation. Depending on the severity and policy settings, a call that violates these policies may be prevented, logged, or an alert raised. By closely examining the runtime behavior of linked tools, this procedure adds an essential layer of operational security.

\subsubsection{Benefits and Challenges}
\textbf{Benefits}: This extension offers highly granular, context-sensitive access control. Users or administrators could review and approve not just broad tool permissions but specific operational policies. It allows for dynamic risk assessment at runtime.

\textbf{Challenges}:
\begin{itemize}
    \item \textbf{Runtime Overhead}: Each policy evaluation adds latency to tool invocation.
    \item \textbf{Complexity}: Designing, managing, and debugging comprehensive Cedar policies or similar requires expertise.
    \item \textbf{Secure Policy Distribution}: Ensuring policies are securely delivered and updated to the PDP is crucial.
    \item \textbf{MCP Server/Client Certification}: To ensure this model is effective, the MCP server and client components that interact with the policy engine must be certified or audited. They must guarantee that all resource accesses are correctly gated by the policy engine and that the contextual information provided to the engine is accurate and complete. Any bypass or manipulation of context would undermine the system.
\end{itemize}

\section{Security Analysis}
The ETDI security model, particularly when enhanced with OAuth 2.0 and further extended with policy-based access control, provides a multi-layered defense against the identified threats. The core ETDI with OAuth effectively counters Tool Poisoning by requiring verifiable cryptographic attestations for tool identity and definitions. Rug Pulls are mitigated by immutable versioning, integrity checks on definitions (which can include API contract hashes), and mandatory re-approval for changes in version or scope. The policy engine layer then adds a dynamic authorization check based on richer context, meaning that even if a tool is generally "approved" and its definition is valid, a specific invocation can be denied if it violates fine-grained contextual policies. This addresses scenarios where a tool might be safe in one context but risky in another.

The efficacy of this combined approach stems from the synergistic application of its core principles:
\begin{itemize}
    \item \textbf{Defense against Tool Impersonation (Tool Poisoning)}:
        \begin{itemize}
            \item \textbf{Cryptographic Authenticity}: OAuth tokens issued by a trusted IdP ensure a tool's claimed identity is backed by a verifiable cryptographic attestation.
            \item \textbf{Provider Verification}: Clients can trust tokens only from specific IdPs and verify the \texttt{iss} (issuer) claim.
            \item \textbf{Binding of Token to Tool Definition}: Custom claims (e.g., \texttt{tool\_id}, \texttt{tool\_version}) link the token to the specific tool definition, preventing token replay for malicious tools.
        \end{itemize}

Figure~\ref{fig:oauth2-prevent-tool-poisoning} demonstrate how OAuth-Enhanced ETDI can be used to prevent tool poisoning attacks.
\begin{figure}[h!]
  \centering
  \includegraphics[width=1\columnwidth]{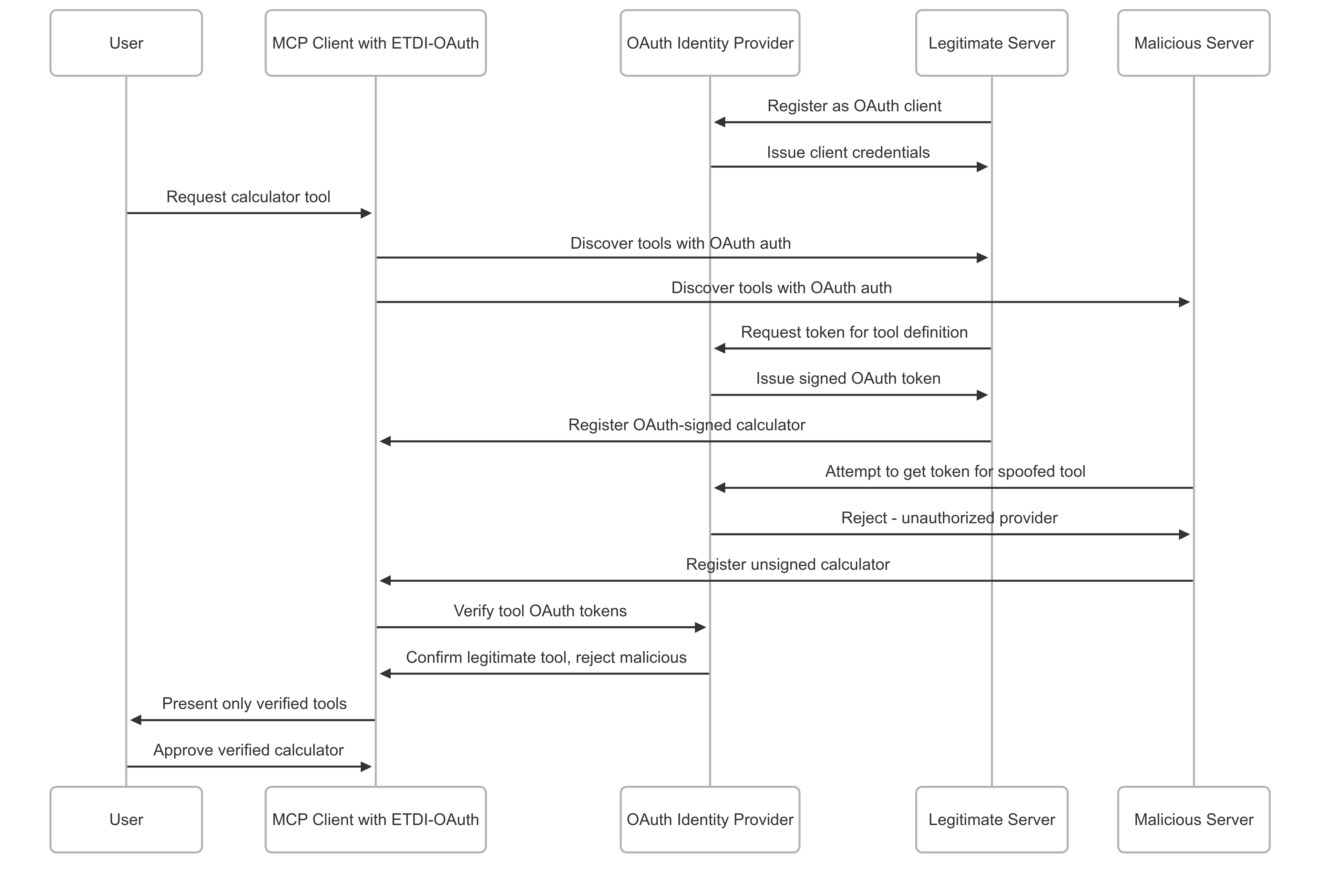} 
  \caption{OAuth-Enhanced ETDI Preventing Tool Poisoning.}
  \label{fig:oauth2-prevent-tool-poisoning}
\end{figure}

    \item \textbf{Defense against Post-Approval Modification (Rug Pulls)}:
        \begin{itemize}
            \item \textbf{Immutable Versioning with Re-Approval}: Changes to security-relevant aspects (permissions/scopes, schema) necessitate a new version number, triggering user re-approval.
            \item \textbf{OAuth Token as a Version-Specific Contract}: Tokens are tied to a specific tool version and scopes. Modifications without a new token may lead to scope mismatches or client detection of definition changes.
            \item \textbf{Scope Adherence}: The client/host can enforce that tools operate only within approved OAuth scopes.
            \item \textbf{Integrity of Stored Approval}: The client stores approved versions and permissions, scrutinizing deviations.
        \end{itemize}

Figure~\ref{fig:oauth2-prevent-rug-pull} demonstrate how OAuth-Enhanced ETDI can be used to prevent rug pull attacks.
\begin{figure}[h!]
  \centering
  \includegraphics[width=1\columnwidth]{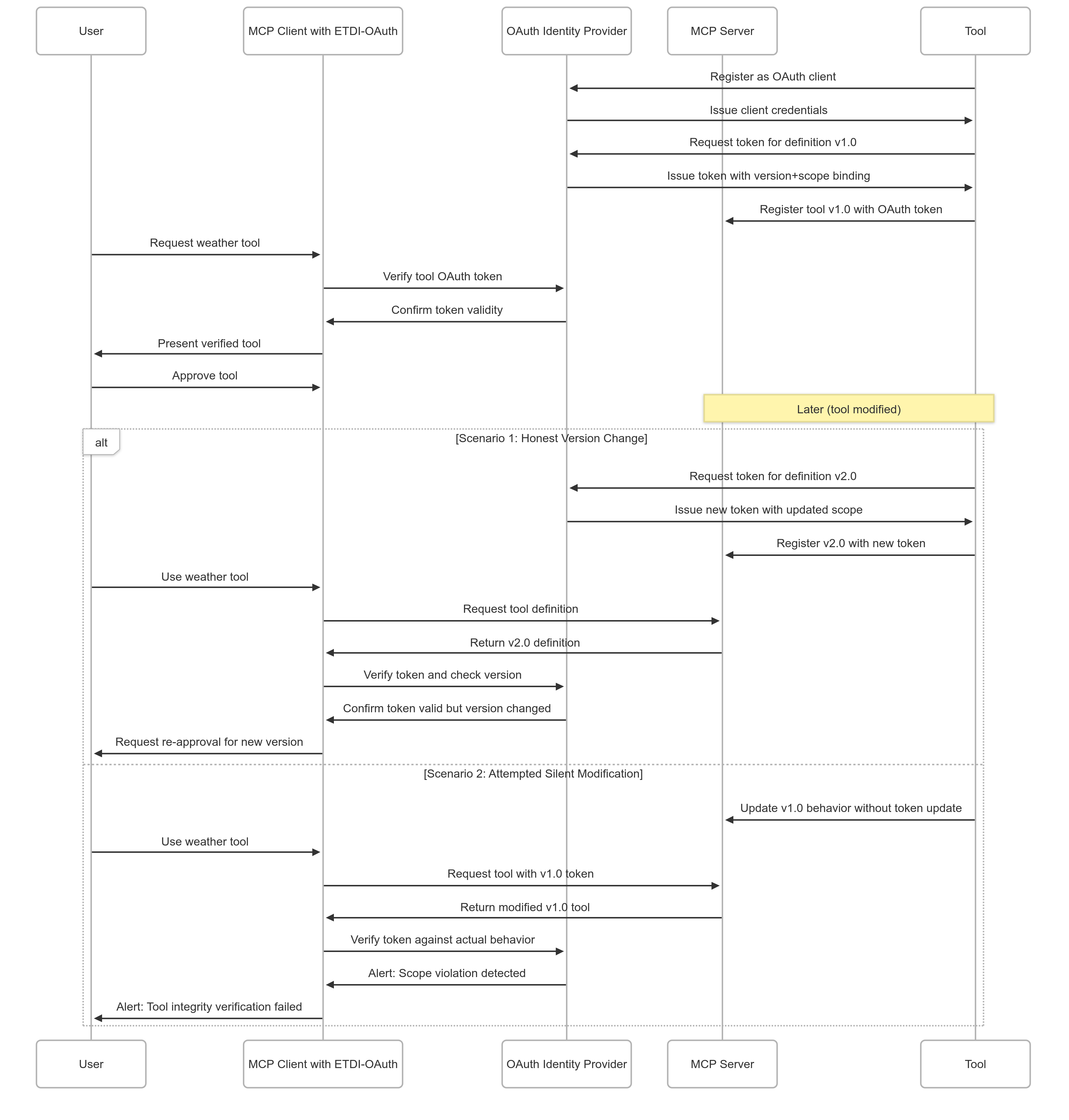} 
  \caption{OAuth-Enhanced ETDI Preventing Rug Pull.}
  \label{fig:oauth2-prevent-rug-pull}
\end{figure}

    \item \textbf{Enhanced Trust and Auditability}:
        \begin{itemize}
            \item \textbf{Centralized Trust Management}: OAuth IdPs centralize tool provider identity and authorization policy management.
            \item \textbf{Standardized Permissions}: OAuth scopes offer standardized permission definitions.
            \item \textbf{Revocation Capabilities}: IdPs support token/credential revocation for swift response to compromises.
        \end{itemize}
\end{itemize}
By requiring cryptographic proof of identity and integrity for each tool version, making changes explicit and subject to re-approval, and adding contextual policy checks, ETDI significantly raises the operational bar for attackers. However, the security of the policy engine itself, the integrity of the policies, and the faithful enforcement by MCP components become critical elements of the trust model.

\section{Discussion}
The introduction of ETDI, its OAuth-enhancement, and the proposed policy-based access control layer address critical security deficiencies in the MCP ecosystem.

\textbf{Trust Assumptions and Scope}: It's important to note that the primary focus of ETDI, as presented, is to address the trustworthiness of *tools* from the perspective of a *trusted user and host application*. This is analogous to the permission model in mobile operating systems (like Android/iOS \cite{felt_android_permissions}), where the user grants permissions to applications (tools) to access platform resources or user data. The user and their primary application environment (Host App, MCP Client) are generally assumed to be non-malicious in this part of the model. The extension to include user/application-specific JWTs being passed to tools (as discussed in Section IV-C) begins to address scenarios where the tool also needs to verify the caller's entitlements, which is common in service-oriented architectures (e.g., checking if a user has paid for a feature).

Several challenges accompany the adoption of ETDI and its extensions:
\begin{itemize}
    \item \textbf{Implementation Complexity}: For tool providers, ETDI-OAuth and policy integration means managing credentials, tokens, scopes, and potentially policy authoring. MCP client/server developers must implement robust validation, policy query logic, and approval workflows.
    \item \textbf{Performance Overhead}: JWT validation and especially remote policy engine calls introduce latency. Caching strategies (for JWKS keys, policy decisions where appropriate) are essential.
    \item \textbf{Key and Policy Management}: Secure management of signing keys (IdP, tool provider, policy administrators) and policy lifecycle is paramount.
    \item \textbf{User Experience (UX)}: Re-approval workflows and explaining policy implications to users must be carefully designed to balance security with usability. Presenting complex Cedar policies for user approval might be challenging; abstractions or summaries would be needed.
    \item \textbf{Ecosystem Adoption}: Broad adoption by tool providers and client applications is necessary for the ecosystem to benefit. Standardization of policy expression and context attributes would be beneficial.
    \item \textbf{Certification of Components}: Particularly for the policy-based extension, certifying that MCP servers and relevant client components correctly and comprehensively enforce policy checks (i.e., they don't bypass the policy engine) is a significant undertaking.
\end{itemize}

Despite these challenges, the security benefits are substantial. This layered approach moves MCP towards a model of verifiable trust and explicit, context-aware authorization.

Future work could explore further decentralization of identity and policy using Verifiable Credentials (VCs) and Decentralized Identifiers (DIDs). Automated analysis of tool behavior against declared permissions and policies could also complement static and dynamic checks.

\section{Literature Review}

Recent studies highlight security vulnerabilities when integrating LLMs with external tools, such as data poisoning, prompt injection, and unauthorized access through compromised APIs \cite{narayanan_llm_security}. Prompt injection attacks, where attackers manipulate model inputs to bypass intended security policies or alter outputs, have been extensively analyzed \cite{jia_gpt_attack}.

Tool poisoning attacks targeting MCP-like protocols have been explored, demonstrating practical examples of malicious tools impersonating trusted entities \cite{willison_mcp_security}. Analyses of MCP’s security gaps \cite{narajala2025enterprise} emphasize the ease of impersonation and lack of cryptographic verification mechanisms in the initial specification \cite{reversinglabs_mcp_risks}.

OAuth 2.0, standardized in RFC 6749 \cite{rfc6749}, provides a framework for delegated authorization, widely adopted for securing API access. Its flexibility allows granular permissions (scopes), reducing over-privilege risks. Comprehensive analyses of OAuth deployments outline common vulnerabilities such as token leakage, session hijacking, and improper scope validation \cite{fett_oauth_attacks}. RFC 7519 (JWT) and RFC 7515 (JWS) further enhance OAuth’s security posture by providing standards for token integrity and authenticity \cite{rfc7519, rfc7515}.

Proper API management and protection, including the use of OAuth to prevent unauthorized API usage, data breaches, and Denial-of-Service (DoS) attacks, are critical. The necessity of cryptographic assurances, identity verification, and robust token lifecycle management is emphasized \cite{schneier_api_security}.

Policy-based access control (PBAC) systems, such as those provided by Amazon Verified Permissions using the Cedar language \cite{amazon_cedar_fse24} and Open Policy Agent (OPA) \cite{opa_docs}, offer dynamic, context-sensitive security enforcement mechanisms. Cedar provides expressiveness and scalability tailored to complex, cloud-native applications. OPA provides a flexible, general-purpose policy evaluation framework adopted across Kubernetes and microservices environments.

PBAC improves security posture in microservice architectures through centralized, fine-grained policy enforcement, minimizing privilege escalation and unauthorized access by embedding policy checks into service logic \cite{oso_microservices_auth}. This literature emphasizes PBAC’s strengths in enforcing granular, context-driven authorization decisions beyond static scopes provided by OAuth.

Immutable and versioned software definitions are foundational security strategies explored in supply-chain security literature. The importance of immutable build artifacts and cryptographically signed components in preventing supply-chain attacks is demonstrated, highlighting parallels to ETDI's approach for tool definitions \cite{torres_arias_supply_chain}.

Docker and Kubernetes ecosystems have adopted immutability and versioning as fundamental principles. The effectiveness of immutability and signature verification in container security parallels ETDI's versioned and signed tool definitions, reducing risks associated with unauthorized modifications \cite{cappos_container_security}.

Cryptographic verification methods, such as digital signatures, are cornerstones of software and API security. Foundational work on RSA encryption \cite{rsa_algorithm} paved the way for widely adopted cryptographic verification practices. Comprehensive coverage of cryptographic methods for identity verification and secure communications underpins ETDI’s use of cryptographic signatures for tool authenticity verification \cite{katz_modern_crypto}.

Exploration of decentralized identities (DIDs) and verifiable credentials (VCs) provides an additional layer of security relevant to ETDI. The benefits of decentralized approaches in identity management, emphasizing improved resilience, privacy, and reduced reliance on central authorities, are extensively documented \cite{allen_did_vc}. These concepts could further extend ETDI, enhancing security and reducing single points of failure.

Usable security, especially in permission systems like ETDI, is critical. Studies on Android permission models conclude that clarity, minimal friction, and explicit consent significantly improve user security behavior and reduce inadvertent risk acceptance \cite{felt_android_permissions}. These insights guide ETDI’s approach in balancing security with practical usability.

\section{Conclusion}
The Enhanced Tool Definition Interface (ETDI), augmented by OAuth 2.0 and a fine-grained policy-based access control layer, provides essential and robust security enhancements for the Model Context Protocol. By systematically incorporating cryptographic identity, immutable versioned definitions (including API contract attestations), explicit permission/scope management, and dynamic contextual policy evaluation, this framework effectively mitigates critical vulnerabilities like Tool Poisoning and Rug Pull attacks. This layered approach fosters a significantly safer and more trustworthy environment for AI applications. Ultimately, these mechanisms aim to bolster user confidence and enable the responsible expansion of LLM-integrated systems by ensuring that tool interactions are based on verifiable trust, explicit consent, and continuously enforced, context-aware policies.

\bibliographystyle{IEEEtran}
\bibliography{references}

\begin{thebibliography}{10}
\providecommand{\url}[1]{#1}
\csname url@samestyle\endcsname
\providecommand{\newblock}{\relax}
\providecommand{\bibinfo}[2]{#2}
\providecommand{\BIBentrySTDinterwordspacing}{\spaceskip=0pt\relax}
\providecommand{\BIBentryALTinterwordstretchfactor}{4}
\providecommand{\BIBentryALTinterwordspacing}{\spaceskip=\fontdimen2\font plus
\BIBentryALTinterwordstretchfactor\fontdimen3\font minus \fontdimen4\font\relax}
\providecommand{\BIBforeignlanguage}[2]{{%
\expandafter\ifx\csname l@#1\endcsname\relax
\typeout{** WARNING: IEEEtran.bst: No hyphenation pattern has been}%
\typeout{** loaded for the language `#1'. Using the pattern for}%
\typeout{** the default language instead.}%
\else
\language=\csname l@#1\endcsname
\fi
#2}}
\providecommand{\BIBdecl}{\relax}
\BIBdecl

\bibitem{mcp_official_spec}
{Model Context Protocol Working Group}, ``{Model Context Protocol Specification},'' [Online], Mar. {2025}, version 2025-03-26. Available: \url{https://modelcontextprotocol.io/specification/2025-03-26/}.

\bibitem{reversinglabs_mcp_risks}
{ReversingLabs Team}, ``{MCP is a powerful new AI coding technology: Understand the risks},'' \emph{ReversingLabs Blog}, 2024, available: \url{https://www.reversinglabs.com/blog/mcp-powerful-ai-coding-risk}.

\bibitem{willison_mcp_security}
S.~Willison, ``{Model Context Protocol has prompt injection security problems},'' \emph{Simon Willison's Weblog}, Apr. 2024, https://simonwillison.net/2025/Apr/9/mcp-prompt-injection.

\bibitem{sarig_agent_threat_model}
D.~Sarig, A.~Maoz, Y.~FOGEL, I.~Harel, and B.~Jimmy, ``{Securing Agentic AI: A Comprehensive Threat Model and Mitigation Framework for Generative AI Agents},'' \emph{arXiv preprint arXiv:2405.03221}, May 2024, available: \url{https://www.researchgate.net/publication/391247531_Securing_Agentic_AI_A_Comprehensive_Threat_Model_and_Mitigation_Framework_for_Generative_AI_Agents}.

\bibitem{rfc6749}
\BIBentryALTinterwordspacing
D.~Hardt, ``The oauth 2.0 authorization framework,'' RFC 6749, 2012. [Online]. Available: \url{https://datatracker.ietf.org/doc/html/rfc6749}
\BIBentrySTDinterwordspacing

\bibitem{rfc7519}
\BIBentryALTinterwordspacing
M.~Jones, J.~Bradley, and N.~Sakimura, ``Json web token (jwt),'' RFC 7519, 2015. [Online]. Available: \url{https://datatracker.ietf.org/doc/html/rfc7519}
\BIBentrySTDinterwordspacing

\bibitem{rfc7515}
\BIBentryALTinterwordspacing
M.~Jones, ``Json web signature (jws),'' RFC 7515, 2015. [Online]. Available: \url{https://tools.ietf.org/html/rfc7515}
\BIBentrySTDinterwordspacing

\bibitem{amazon_cedar_fse24}
\BIBentryALTinterwordspacing
S.~Schuster, M.~Backes, S.~Stoller, P.~H.~S. Torr, K.~SefidMahan, B.~K{\"o}pf, J.~Aldrich, S.~Chong, G.~Roşu, A.~Rybalchenko \emph{et~al.}, ``{How We Built Cedar: A Verification-Guided Approach to Authorization},'' in \emph{Companion Proceedings of the ACM on Programming Languages (FSE Companion)}, Jul. 2024, note: Based on search result for FSE '24. Actual publication details might vary slightly or refer to the general availability and open-sourcing of Cedar. For the purpose of this example, using the found paper title. A direct AWS whitepaper might also be suitable. [Online]. Available: \url{https://assets.amazon.science/d3/86/99db1aa142ffb6981d86dc849e4c/how-we-built-cedar-a-verification-guided-approach.pdf}
\BIBentrySTDinterwordspacing

\bibitem{opa_docs}
{Open Policy Agent Team}, ``{Introduction to Open Policy Agent},'' available: \url{https://www.openpolicyagent.org/docs/latest/}.

\bibitem{Narajala2025ToolSquatting}
\BIBentryALTinterwordspacing
V.~S. Narajala, K.~Huang, and I.~Habler, ``Securing genai multi-agent systems against tool squatting: A zero trust registry-based approach,'' \emph{arXiv preprint arXiv:2504.19951}, 2025. [Online]. Available: \url{https://arxiv.org/abs/2504.19951}
\BIBentrySTDinterwordspacing

\bibitem{oso_microservices_auth}
{Oso Team}, ``{Best Practices for Authorization in Microservices},'' \emph{Oso Blog}, 2023, available: \url{https://www.osohq.com/post/microservices-authorization-patterns}.

\bibitem{felt_android_permissions}
A.~P. Felt, E.~Chin, S.~Hanna, D.~Song, and D.~Wagner, ``{Android Permissions: User Attention, Comprehension, and Behavior},'' in \emph{Proceedings of the Eighth ACM Symposium on Usable Privacy and Security (SOUPS '12)}.\hskip 1em plus 0.5em minus 0.4em\relax ACM, 2012, pp. 3:1--3:14, also available as UC Berkeley EECS Technical Report EECS-2012-26.

\bibitem{narayanan_llm_security}
\BIBentryALTinterwordspacing
A.~Narayanan and et~al., ``Persistent attacks on large language models beyond user injection,'' \emph{arXiv preprint arXiv:2505.06493}, 2025. [Online]. Available: \url{https://arxiv.org/abs/2505.06493}
\BIBentrySTDinterwordspacing

\bibitem{jia_gpt_attack}
\BIBentryALTinterwordspacing
J.~Jia and et~al., ``Formalizing and benchmarking prompt injection attacks and defenses,'' \emph{USENIX Security Symposium}, 2024. [Online]. Available: \url{https://www.usenix.org/system/files/usenixsecurity24-liu-yupei.pdf}
\BIBentrySTDinterwordspacing

\bibitem{narajala2025enterprise}
\BIBentryALTinterwordspacing
V.~S. Narajala and I.~Habler, ``{Enterprise-Grade Security for the Model Context Protocol (MCP): Frameworks and Mitigation Strategies},'' \emph{arXiv preprint arXiv:2504.08623}, 2025. [Online]. Available: \url{https://arxiv.org/abs/2504.08623}
\BIBentrySTDinterwordspacing

\bibitem{fett_oauth_attacks}
\BIBentryALTinterwordspacing
D.~Fett, R.~Kuesters, and G.~Schmitz, ``A comprehensive formal security analysis of oauth 2.0,'' \emph{Proceedings of the 2016 ACM SIGSAC Conference on Computer and Communications Security}, pp. 1204--1215, 2016. [Online]. Available: \url{https://publ.sec.uni-stuttgart.de/fettkuestersschmitz-ccs-2016.pdf}
\BIBentrySTDinterwordspacing

\bibitem{schneier_api_security}
\BIBentryALTinterwordspacing
B.~Schneier, ``Reports on api security,'' 2025. [Online]. Available: \url{https://www.schneier.com/tag/reports/}
\BIBentrySTDinterwordspacing

\bibitem{torres_arias_supply_chain}
\BIBentryALTinterwordspacing
S.~Torres-Arias and et~al., ``A new tool wants to save open source from supply chain attacks,'' \emph{WIRED}, 2021. [Online]. Available: \url{https://www.wired.com/story/sigstore-open-source-supply-chain-code-signing/}
\BIBentrySTDinterwordspacing

\bibitem{cappos_container_security}
\BIBentryALTinterwordspacing
ReversingLabs, ``7 container security best practices,'' 2025. [Online]. Available: \url{https://www.reversinglabs.com/blog/7-container-security-best-practices}
\BIBentrySTDinterwordspacing

\bibitem{rsa_algorithm}
R.~L. Rivest, A.~Shamir, and L.~Adleman, ``A method for obtaining digital signatures and public-key cryptosystems,'' \emph{Communications of the ACM}, vol.~21, no.~2, pp. 120--126, 1978.

\bibitem{katz_modern_crypto}
J.~Katz and Y.~Lindell, \emph{Introduction to Modern Cryptography}.\hskip 1em plus 0.5em minus 0.4em\relax Chapman and Hall/CRC, 2014.

\bibitem{allen_did_vc}
\BIBentryALTinterwordspacing
C.~Allen, ``Decentralized identifiers (dids) v1.0,'' 2021. [Online]. Available: \url{https://www.w3.org/TR/did-core/}
\BIBentrySTDinterwordspacing

\end{thebibliography}

\end{document}